\documentclass[a4paper]{article}

\usepackage{icrc2013}

\title{Developments for coating, testing, and aligning Cherenkov Telescope Array mirrors in T\"ubingen}

\shorttitle{Bonardi {\footnotesize et al.} Study for CTA mirrors}

\authors{A.~Bonardi$^{1}$, J.~Dick$^{1}$, E.~Kendziorra$^{1}$, G.~P\"uhlhofer$^{1}$, A.~Santangelo$^{1}$
 for the CTA consortium
}

\afiliations{
$^1$ Institut f\"ur Astronomie und Astrophysik T\"ubingen\\
}

\email{antonio.bonardi@uni-tuebingen.de}

\abstract{The Cherenkov Telescope Array (CTA) is the next generation very-high energy gamma-ray air-shower Cherenkov observatory. CTA will consist of many segmented-mirror telescopes of three different diameters, placed in two arrays, one in the Northern hemisphere and one in the South,
thus covering the whole sky. The total number of mirror tiles will be of the order of 10000, corresponding to a reflective area of $\sim 10^{4}$~m$^{2}$. The Institute for Astronomy and Astrophysics in T\"ubingen is developing procedures to coat glass-substrate-based mirror tiles, is participating to the CTA mirror prototype testing, and is prototyping Active Mirror Control alignment mechanics, electronics and software.
We will present the current status of our work and plans for future developments.}


\begin{document}
\maketitle

\section{Introduction}
The Cherenkov Telescope Array (CTA) consortium aims to deploy two arrays of Imaging Atmospheric Cherenkov Telescopes (IACT), one in the Southern and one in the Northern hemisphere, each of them covering an area of $\sim$1-3 km$^{2}$. At least three different telescope classes are foreseen: a large size telescope (LST) of $\sim$24 m dish diameter, a medium size telescope (MST) of $\sim$12 m, and a small size telescope (SST) of $\sim$4-7 m \cite{bib:CTA_concept}. The reflective surface of each telescope primary mirror will consist of hexagonal mirror facets of $\sim$0.5-2 m$^{2}$ area, according to the telescope type. The baseline design for the LST and MST is a single-reflector telescope (parabolic dish for LST, spherical dish for MST) with the camera in the primary focus. In the SST case both a single-reflector design with a spherical dish and a dual-reflector design with a Schwarzschild-Couder configuration are in parallel development.    


The total reflective mirror surface of the entire CTA is huge ($\sim10^{4} $m$^{2}$), hence  several thousand
mirror segments will be required and high efficiency during testing and alignment of the segments is necessary. High durability of the mirror reflective surface is equally necessary to avoid the need for frequent re-coating. In general, segments should remain sufficiently reflective for at least 10 years. In this paper, activities by the Institute for Astronomy and Astrophysics in T\"ubingen (IAAT) conducted in the field of mirror alignment, mirror testing, and mirror coating are reported. The work is performed in the framework of the CTA mirror work package.

\section{Mirror coating}
Current mirror facet specifications demand that the reflected light should largely be contained in a 1 mrad diameter area, the reflectance in the range 300 nm $\leq \lambda \leq$ 600 nm (hereafter WR$_{300-600}$) should be $\geq 80\%$, and facets should be robust against aging for several years \cite{bib:CTA_concept}. The goal is of course to provide coatings with the highest possible reflectance and the longest economically possible lifetime. 

Currently, various designs for mirror facets are under study, including aluminum or carbon fiber honeycomb structures. The focus of the study presented here is mirror types where a reflective layer on top of a glass substrate is needed. Such coating consists of a thin (100-2000 nm) single or multiple reflective layer and, if needed, a protective overcoating.

\subsection{Mirror coating study}
\begin{figure}[!htb]
  \centering
  \includegraphics[width=0.4\textwidth]{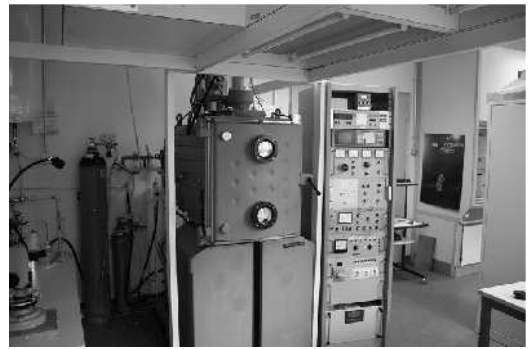}
  \caption[fig01]{\footnotesize Coating chamber of the IAAT.}
\label{fig01}
\end{figure}

The aim of our study is to obtain a mirror coating solution with reflectance $> 90 \%$ (measured close to the mirror surface, light incidence angle $\theta=0$) for the entire WR$_{300-600}$, with high durability. At the same time, it should be as simple as possible in order to be suitable for mass production. We currently use the McLeod \cite{bib:McLeod} simulation software for optimizing our coating designs, which are then realized on small glass substrates in the IAAT coating chamber (shown in Fig.~\ref{fig01}).

The IAAT coating chamber is equipped with:
\begin{itemize}
\item a rotary pump for vacuum production ($10^{-6}-10^{-7}$~mbar);
\item a nitrogen flooding system for humidity removal and used for opening the chamber;
\item an oxygen flooding system for impurity removal and reactive oxide deposition;
\item a water warming/cooling system (15-65 $^{\textrm o}$C);
\item one resistor crucible for metal evaporation;
\item a 4-fold electron beam crucible for storing and evaporating up to four different dielectric materials;
\item a 6-fold Provac QSP 650 quartz micro-balance for measuring the deposited layer thickness of up to six different materials.
\end{itemize}

For the CTA mirror coating, three main solutions are followed (for more details see \cite{bib:afoerster}):
\begin{itemize}
 \item[1.] one metallic reflective layer plus a very robust protective overcoating;
 \item[2.] one metallic reflective layer plus a multilayer interferometric overcoating for both protection and reflectance improvement;
 \item[3.] a purely dielectric multilayer interferometric coating.
\end{itemize}

The first solution usually consists of an Al layer with a SiO$_{2}$ protective layer. Such a solution is widely in use for current IACT mirror facets because of its good performance, ease of manufacture, and low cost. Its biggest drawback is the long term reliability achieved so far which is most likely inadequate for CTA scales.

The second solution is an intermediate solution. Enhancing a simple reflective layer (usually Al) with a protective coating not only provides a protection to the metallic layer but also improves the reflectance in the WR$_{300-600}$.

The third solution is based on the reflection of the light when passing from a material to another with different refractive index. The reflection is described by Fresnel's law 

\begin{eqnarray*}
 R = \left(\frac{n_{1}-n_{2}}{n_{1}+n_{2}}\right)^{2}
\end{eqnarray*}

where $R$ is the reflection coefficient, and $n_{1}$ and $n_{2}$ are the refractive indexes of the first and second material.\\
For a light ray passing from one layer to another, the reflectance also depends on its wavelength because of self-interference effects. The maximum reflectance value is obtained for

\begin{eqnarray*}
 \lambda = \frac{4\cdot n_{1} \cdot L_{1}}{\sin(\theta)}
\end{eqnarray*}

where $\lambda$ is the light wavelength in vacuum, $L_{1}$ and $n_{1}$ are respectively the thickness and the refractive index of the first layer, and $\theta$ is the incident angle. 
By alternating many dielectric layers of different thickness and refractive index, it is possible to achieve both a very high reflectance inside the WR$_{300-600}$ and a low reflectance outside of it. 
In this way, the Cherenkov light collection will be maximized, and at the same time the pollution by the Night Sky Background (NSB) -- mostly at large wavelength -- will be minimized. 
Furthermore, since no metallic layer is used, deterioration due, for example, to oxidation will not occur. On the other hand, a large number  of layers ($> 25$) is required. This could result in layer deterioration in the field due to the thermally induced stresses between different layers. 
Furthermore, a higher humidity condensation rate has been recently observed compared to metallic or intermediate coatings \cite{bib:pchadwick}.\\
Since the last solution is so far the least explored one (at least concerning IACTs), we have focused our research on a reliable and easy way to produce an interferometric coating design. \\

\subsection{Present mirror-coating results}
\begin{figure}[!htb]
  \centering
  \includegraphics[width=0.4\textwidth]{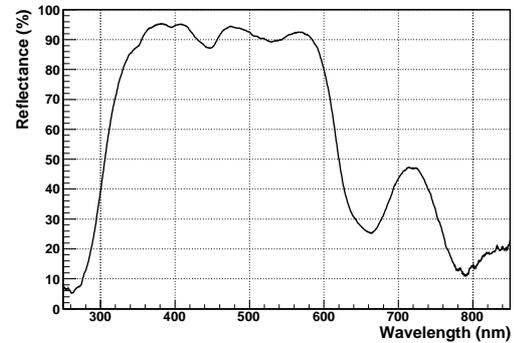}
\caption[fig02]{\footnotesize Measured reflectance vs. wavelength for light normally incident on a 24 layer TiO$_{2}$-SiO$_{2}$ coated glass substrate. During TiO$_{2}$ deposition the oxygen pressure in the chamber is P(O$_{2}$)=10$^{-4}$ mbar.}
\label{fig02}
\end{figure}

The main difficulty we encountered in our work was that many high index materials, which are transparent in the WR$_{300-600}$ (e.g. Ta$_{2}$O$_{5}$), need heating up to about $350~{^{\textrm o}}$C during or after the evaporation and deposition. While such treatment is not problematic for glass mirrors, no honeycomb structure mirrors can endure temperatures above $\sim80~{^{\textrm o}}$C without suffering irreversible damage. 
By following the experience of Duyar et al. \cite{bib:duyar}, we investigated the possibility of achieving a reflection $R(300<\lambda <600)$ greater than $90\%$ with a 24-layer dielectric coating design made of TiO$_{2}$ and SiO$_{2}$. The TiO$_{2}$ was deposited through Ti$_{3}$O$_{5}$ reactive evaporation in an O$_{2}$ atmosphere with pressure P(O$_{2}$)=$10^{-4}$ mbar and without any substrate thermal heating. Unfortunately, the transparency of the resulting TiO$_{2}$ layers is too low below 350 nm, resulting in a very poor reflectance for wavelength encompassed between 300 nm and 350 nm as shown in Fig.~\ref{fig02}.

\begin{figure}[!htb]
  \centering
  \includegraphics[width=0.4\textwidth]{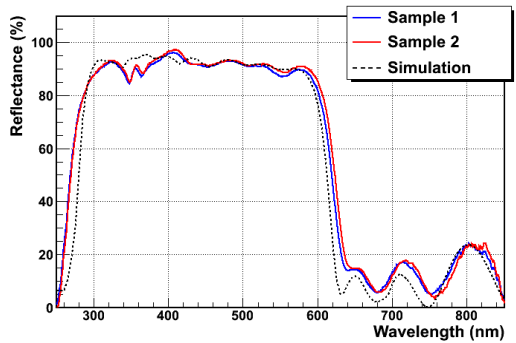}
\caption[fig03]{\footnotesize Measured reflectance vs. wavelength for light normally incident on a 28-layer TiO$_{2}$-SiO$_{2}$-HfO$_{2}$ coated glass substrate, real data (blue and red solid line) and simulation (black dashed line). During TiO$_{2}$ deposition the oxygen pressure in the chamber is P(O$_{2}$)=10$^{-4}$ mbar.}
\label{fig03}
\end{figure}

Therefore, we modified the previous design by adding 2 layers of SiO$_{2}$-HfO$_{2}$ over the topmost layer, resulting in 28 overall layers, in order to enhance the reflectance in the 300-350 nm region too. The reflectance spectrum of two samples of the 28-layer TiO$_{2}$-SiO$_{2}$-HfO$_{2}$ design produced in the IAAT coating chamber is shown in Fig.~\ref{fig03} together with the spectrum obtained from simulations.

Fig.~\ref{fig03} shows that our samples are already well within our target, i.e. $R(WR_{300-600})>90\%$, and very low reflectance for $\lambda > 600$ nm, with no substrate heating. At the present time study is ongoing for both improving the reflectance further in the WR$_{300-600}$ region and facing the problem of humidity condensation for dielectric-coated mirrors as discussed above.\\

\section{Mirror reflectance and psf testing}

\begin{figure}[!htb]
  \centering
  \includegraphics[width=0.4\textwidth]{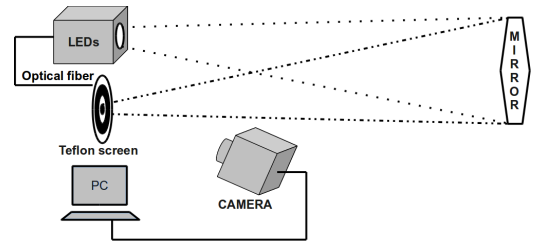}
\caption[fig04]{\footnotesize The new 2f mirror test setup based at IAAT.}
\label{fig04}
\end{figure}

The leading idea of any 2f mirror test setup is to place a point-like light source and a target screen at two times the mirror nominal focal distance~(2f), and to observe the so-obtained light spot. In such way, both the mirror PSF and its reflectance on the focal plane can be determined. 
A 2f mirror test stand developed by the MPI-K Heidelberg and housed in the IAAT main building was used for checking the optical qualities (i.e. reflectance and spot size) of all the mirror facets for the H.E.S.S. II telescope \cite{bib:icrc2011}. Starting from that experience, IAAT developed a new 2f mirror test setup able to test all the different mirror facet types which will be used for the three CTA single-reflector telescope classes (SST, MST, LST).  
The setup is shown in Fig.~\ref{fig04} and summarized in the following:

\begin{itemize}
\item two LEDs, a 310 nm UV LED and a 4-chip multi-wavelength one, are used for lighting up the tested mirror with monochromatic light, so that all the WR$_{300-600}$ region is covered; 
\item the mirror is placed on a movable support which is placed at twice the mirror focal distance from the LEDs;
\item the reflected light converges on a Teflon screen, placed close to the LEDs;
\item a small fraction of the light emitted by the LEDs is directly conveyed to the Teflon screen by optical fibers;
\item the reflected light spot at the different wavelengths is observed with an ALTA U47 CCD camera and, thus, the mirror PSF (Point Spread Function) is measured;
\item for each wavelength the total reflectance value is obtained by comparing the intensity of the light spot produced with light channeled by the optical fiber to the screen.
\end{itemize}

Thanks to the fast ALTA U47 read-out and to the ease of controlling the different monochromatic light sources, the new IAAT mirror test setup can perform a complete measurement of the mirror PSF and reflectance on the focal plane in few minutes, plus the time required for preparing the mirror on the test support. Thus, we expect that the new IAAT mirror test setup could reasonably measure up to to 6 CTA mirrors per hour. 

\section{Mirror alignment system}

The large number of mirror facets, especially for the MST and LST types, demand motorized control of the mirror actuators, even if the mirror dish is stiff enough that only an initial alignment after mirror mounting or mirror exchange is needed. In the current baseline design for the MST, the dish is indeed expected to be stiff
enough that an active alignment (i.e. frequent realignment during telescope observation time) will not be needed. On the other hand, the LST structure will most likely require such active alignment. While the demand on mirror actuators is certainly lower for initial alignment procedures, nevertheless options which might be suited for both types of alignment are being pursued.

\begin{figure}[!htb]
  \centering
  \includegraphics[width=0.4\textwidth]{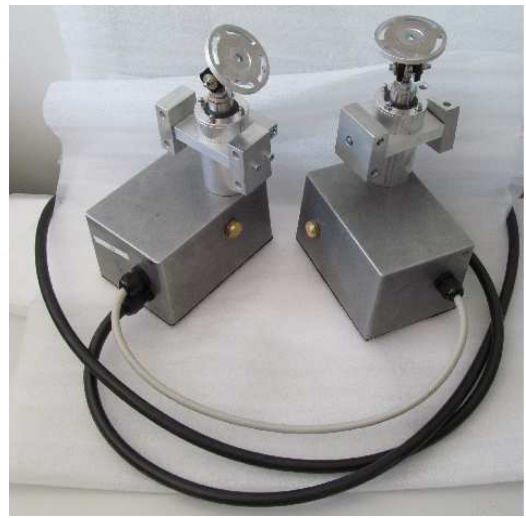}
\caption[fig05]{\footnotesize A pair of the new pre-series IAAT actuators.}
\label{fig05}
\end{figure}

The design which the IAAT is currently developing and testing originates in the actuator mechanics developed for the large, 27 m telescope of H.E.S.S. phase II \cite{bib:hess_amc}. Since 2012 IAAT has collaborated with Buck Engineering Consulting GmbH (with the CTA IRD financial support) to improve the actuator mechanics design with the main goal of optimizing the cost-performance balance. Recently 32 new pre-series actuators (see Fig.~\ref{fig05}) have been produced and installed on the MST prototype built by DESY in Adlershof-Berlin, and are currently under test. 

\begin{figure}[!htb]
  \centering
  \includegraphics[width=0.4\textwidth]{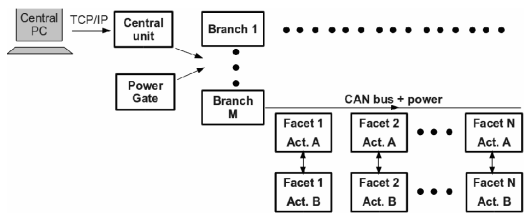}
\caption[fig06]{\footnotesize Scheme of the actuator control system developed for CTA telescopes.}
\label{fig06}
\end{figure}

As in the case of the H.E.S.S. phase II actuators, mirror facets are supported by two motor driven actuators and one freely-rotating bolt which is fixed to the telescope structure. In the current MST design, the actuators are connected to a supporting triangle, which is fixed to the structure of the telescope dish by two clamps. One actuator is tightly anchored to the supporting triangle so as not to allow any tilting movement; the second one is equipped with a pivot, allowing free rotation along a tilting angle. The motor of each actuator is housed in a watertight box vented by a sintered bronze valve. The actuator step length is now 5 $\mu$m and the total excursion range is $\sim 45$ mm. Furthermore, as in the case of the H.E.S.S. phase II actuators, the external fixing clamp allows the actuator to shift its position with reference to the telescope dish structure by about 5 cm.\\ 
Compared to the electronics which is used to control the actuators at H.E.S.S. II \cite{bib:hess_amc}, a new architecture based on the Controller Area Network (CAN) interface \cite{bib:CAN} has been developed and employed. For each mirror there is a electronic control board, composed of a driver for each motor and a micro-controller (Atmel AT90CANxx), which is placed in one of the two actuator boxes. The other actuator is connected to the control electronics by an external cable. For each mirror, the two actuators plus the electronic control board form a Mirror Control Unit (MCU). The MCUs are serially connected to a central unit. Each such central unit serves one telescope, and can be placed in the center of the dish. The connection is realized with a four-wire shielded cable: two wires are used for transmitting the CAN signal, the other two for the power supply. A limitation of the chain length is not imposed by the CAN interface, but by the Ohmic power loss of the cables transmitting the motor current. As an alternative, more expensive cables could be used, but we found the best solution consists of a distribution of the MCUs along different CAN chains, each of them serving 7-8 MCUs. The central unit, an embedded PC working as a TCP/IP to CAN gateway, is controlled by a controlling PC. The control PC could be a Client or Server and be placed in the telescope tower or in the central control room, depending on the architecture of the telescope control system. A power gate for the AMC DC power supply is also present in the telescope tower.\\
The scheme of the actuator control system is shown in Fig.~\ref{fig06}.

While in the H.E.S.S. II control frame only one mirror facet can be aligned at a time \cite{bib:hess_amc}, with our new frame it is possible to align a large number of mirror facets at the same time. The only limitation is the AMC power consumption.\\
Furthermore, such a control frame is extremely simple and robust as evidenced by our laboratory and external tests. During different tests, we transmitted various millions of CAN commands without experiencing any transmission failure. It is also extremely flexible and can be easily adapted to any of the CTA telescope types requiring motorized actuators.

\vspace*{0.5cm}
\footnotesize{{\bf Acknowledgment:}{This work has been partially funded by the BMBF/PT-DESY, grants 05A11VT1 and 05A10VTA. We gratefully acknowledge support from the agencies and organizations listed in this page: http://www.cta-observatory.org/?q=node/22 .}}

\end{document}